\documentclass[a4paper,11pt]{article}
\usepackage[latin1]{inputenc}
\usepackage[english]{babel}
\usepackage{graphicx}
\begin{document}
\title{Ten scenarios from early radiation to late time acceleration with a minimally coupled dark energy}
\author{Stéphane Fay\footnote{steph.fay@gmail.com}\\
Palais de la Découverte\\
Astronomy Department\\
Avenue Franklin Roosevelt\\
75008 Paris\\
France}
\maketitle
\begin{abstract}
We consider General Relativity with matter, radiation and a minimally coupled dark energy defined by an equation of state $w$. Using dynamical system method, we find the equilibrium points of such a theory assuming an expanding Universe and a positive dark energy density. Two of these points correspond to classical radiation and matter dominated epochs for the Universe. For the other points, dark energy mimics matter, radiation or accelerates Universe expansion. We then look for possible sequences of epochs describing a Universe starting with some radiation dominated epoch(s) (mimicked or not by dark energy), then matter dominated epoch(s) (mimicked or not by dark energy) and ending with an accelerated expansion. We find ten sequences able to follow this Universe history without singular behaviour of $w$ at some saddle points. Most of them are new in dark energy literature. To get more than these ten sequences, $w$ has to be singular at some specific saddle equilibrium points. This is an unusual mathematical property of the equation of state in dark energy literature, whose physical consequences tend to be discarded by observations. This thus distinguishes the ten above sequences from an infinity of ways to describe Universe expansion.
\end{abstract}
\section{Introduction}\label{s0}
An important problem of today cosmology is to choose between an infinite number of theories which ones are able to explain observations. Most of time, this is done by comparing a theory or a whole family of theories with some observations as supernovae, CMB, weak lensing and so on. Another basic, and at first glance obvious question to validate a cosmological theory, is also to check if it is able to reproduce a Universe history starting by a radiation dominated epoch, followed by a matter dominated epoch and then an accelerated expansion epoch. For instance it has been claimed that this is not possible for $f(R)$ theories in the metric formalism\cite{AmGa07} but if they respect some special conditions, a contested result in \cite{CaTr08}. In this paper, we examine how General Relativity with matter, radiation and a minimally coupled dark energy defined by an equation of state $w$ is able to reproduce the above Universe history whose $\Lambda$CDM description is one way among many others.
\\
To reach this aim, we apply dynamical system method to the equations of General Relativity, rewritten with a set of normalised variables. We get the equilibrium points of this equations system and give some general (since we will keep $w$ unspecified) considerations about their stability. Two of these points represent usual radiation and matter dominated epochs. For the other points, dark energy mimics matter or radiation but for one which corresponds to an accelerated expansion of the Universe. Then we try to determine all the possible trajectories in the phase space describing sequences of epochs starting by some dominated radiation epochs (mimicked or not by dark energy), then some matter dominated epochs (mimicked or not by dark energy) and ending with an accelerated expansion epoch. We find ten of these sequences when $w$ is not singular at some saddle points. Most of them are new in the dark energy literature  like the ones when dark energy mimics radiation. To get more than these ten sequences, a necessary but not sufficient condition is that $w$ be singular at some saddle points. Physically, since a trajectory in the phase space never pass through a saddle point but close to it, it means that $w$ should have pronounced bumps or/and dips, a property of $w$ that tends to be discarded by observations\cite{Bie13}\cite{Mor09}.\\
These results may be seen as a classification of dark energy theories in presence of matter and radiation, based on dynamical system. There are already several dark energy classifications in the literature. In \cite{BarGua06}, dark energy theories are classified in a plane of observables that corresponds to the parameterisation of a variable equation of state, $w(a)=w_0+w_a(1-a)$, where $a$ is the scale factor of the universe. A similar study has been made in \cite{Chen09}. Another way of classifying dark energy models is using statefinder parameters\cite{Eva05}. In \cite{SzyAle07}, the class of all dark energy models is considered as a multiverse and classified using the notion of Banach space. The classification we get in this paper is different from these ones. It is based on the notion of matter, radiation and dark energy equilibrium points but it does not distinguish between a quintessence or a ghost model for instance.\\
The plan of this paper is as follows. In the second section, we look for all possible equilibrium points of the dynamical system defined by equations of General Relativity with matter, radiation and a minimally coupled dark energy. We give some general considerations about their stability. In the third section, we look for some possible sequences of epochs describing Universe first dominated by radiation, then matter and then dark energy accelerating expansion. We find $10$ of these sequences when $w$ is not singular at some saddle equilibrium points. We discuss our results in the last section. We explain why $w$ should be singular at some saddle equilibrium points to reproduce an infinity of sequences and not only the $10$ of the third section. This property of $w$ could be in tension with the data, thus distinguishing the $10$ sequences we found in the third section among all the possible ones.
\section{Field equations}\label{s1}
Field equations of General Relativity with a dark energy in a flat FLRW Universe write
$$
H^2=\frac{k}{3}(\rho_m+\rho_r+\rho)
$$
$$
\frac{dH}{dt}=\frac{k}{6}\left[-3\rho_m-4\rho_r-3(w+1)\rho\right]
$$
$\rho_m$ is the matter density. From conservation of the stress-energy tensor for $\rho_m$ we derive that $\rho_m=\rho_{m0}(1+z)^3$, with $z$ the redshift and $\rho_{m0}$ the matter density today. $\rho_r$ is the radiation density. From conservation of the stress-energy tensor for $\rho_r$ we derive that $\rho_r=\rho_{r0}(1+z)^4$ with $\rho_{r0}$ the radiation density today. $\rho$ is the dark energy density with equation of state $p/\rho=w$, $p$ being the dark energy pressure. Conservation of the stress-energy tensor for $\rho$ writes
\begin{equation}\label{ecDE}
\frac{d\rho}{dt}=-3H(w+1)\rho
\end{equation} 
Then we define
\begin{equation}\label{y1}
y_1=\frac{k}{3}\frac{\rho_m}{H^2}
\end{equation}
\begin{equation}\label{y2}
y_2=\frac{k}{3}\frac{\rho_r}{H^2}
\end{equation}
\begin{equation}\label{y3}
y_3=\frac{k}{3}\frac{\rho}{H^2}
\end{equation}
With the constraint equation
$$
1=y1+y2+y3
$$
We are going to make two assumptions:
\begin{enumerate}
\item $H>0$, i.e. Universe is expanding in agreement with the data.
\item $\rho>0$, i.e. dark energy density is positive. Negative energy density generally belongs to the realm of quantum cosmology that we will not consider in this work.
\end{enumerate}
Since $\rho_m>0$, $\rho_r>0$ and we assume $\rho>0$, the above equation implies that $0<y_i<1$. Differentiating $y_1$ and $y_2$ and considering field equations and energy density conservation of $\rho_m$, $\rho_r$ and $\rho$, we are left with the following dynamical system
\begin{equation}\label{y1p}
y_1'=-3y_1-2y_1\left[-\frac{3}{2}y_1-2y_2-\frac{3}{2}(1+w)(1-y_1-y_2)\right]
\end{equation}
\begin{equation}\label{y2p}
y_2'=-4y_2-2y_2\left[-\frac{3}{2}y_1-2y_2-\frac{3}{2}(1+w)(1-y_1-y_2)\right]
\end{equation}
A prime stands for a derivative with respect to $N=\ln a$, with $a$ the scale factor. Since we assume that Universe is expanding, this allows to consider $N$ as a time variable, increasing with proper time $t$. We derive an equation for $w$ from (\ref{y1p}-\ref{y2p}):
\begin{equation}\label{w}
w=\frac{1}{3(-1+y_1+y_2)}\left(\frac{y_2\frac{dy_1}{dy_2}}{y_1-\frac{dy_1}{dy_2}y_2}+y_2\right)
\end{equation}
The equations system (\ref{y1p}-\ref{y2p}) is autonomous since $w$ can always be written as a function of $y_1$ and $y_2$. This last statement needs explanations. It does not imply any coupling between matter, radiation and dark energy. Basically, it is due to the fact that $w$ can always be considered as a function of the redshift $z$. Then, since from (\ref{y1}-\ref{y2}), we get $z=(\rho_{m0}/\rho_{r0} y_2/y_1-1)$, we can write, at least formally, $w=w(y_2/y_1)$. Finding the explicit form of $w(y_1,y_2)$ is obvious when the equation of state $w$ is given as a function of $z$. For instance the linear equation of state\cite{CooHut99} is $w=w_0+w_1 z$, and thus $w(y_1,y_2)=w(y_2/y_1)=w_0+w_1(\rho_{m0}/\rho_{r0} y_2/y_1-1)$. But $w$ can also take more general forms than that of a function of $y_2/y_1$. For instance, this is the case when one looks for $w$ mimicking the Dvali-Gabadadze-Porrati\cite{DvaGabPor00} theory. Its modified Friedmann equation is $H^2\pm H/r_c = k/(6 r_c)(\rho_m+\rho_r)$ with $r_c$ the crossover scale\cite{LomHu09}. If we put the $\pm H/r_c$ term on the right-hand side of the Einstein equations, defining the constant $K$ by $k=2 r_c K$, this theory is mimicked by a dark energy theory with $\rho = +2(r_c K)^{-1}H$ (we take the "+" branch of the DGP theory since we assume $\rho>0$). Using equation (\ref{ecDE}), one shows that the corresponding $w$ is then $w=(y_2-3)/\left[3(1+y_1+y_2)\right]=w(y_1,y_2)$\footnote{Even for the DGP theory, it is always possible to consider $w$ as a function of the redshift and to write formally $w=w(y_2/y_1)$. In this case the mathematical relation $w(y_2/y_1)=w(y_1, y_2)$ expresses the fact that the two independent functions $y_1$ and $y_2$ both depend on the redshift $z$. Indeed, since $y_1=y_1(z)$ and $y_2=y_2(z)$, it is always possible to write formally $y_1(y_2)$ or equivalently the relation $w(y_2/y_1)=w(y_1, y_2)$.}. Hence, in all this paper, we will write $w$ as $w(y_1,y_2)$.\\\\
Summarising, we assume that Universe is expanding and dark energy density is positive. The field equations (\ref{y1p}-\ref{y2p}) reduce to a system of two equations for $y_1$ and $y_2$ with $w(y_1,y_2)$ and can be studied with usual dynamical system method. Stability of the equilibrium points depends on the signs of the eigenvalues
\begin{equation}\label{sta}
\lambda_\pm=\frac{1}{2}\left[A\pm\sqrt{(A-A_+)(A-A_-)}\right]
\end{equation}
with
$$
A=-1+3y_ 2+3(2-3y_1-3y_2)w-3(-1+y_1+y_2)(y_2 w_{y_2}+y_1 w_{y_1})
$$
$$
A_\pm=2\left[y_2+3(1-y_1-y_2)w\pm\sqrt{3}\sqrt{y_1(w-(1-y_1-y_2)w_{y_1})}\right]
$$
and $w_{y_i}$, the derivative of $w(y_1,y_2)$ with respect to $y_i$. 
\subsection{Equilibrium points}\label{s11}
Now, we look for the equilibrium points of the system (\ref{y1p}-\ref{y2p}) whatever $w$.\\
First we show that there is no equilibrium point with both $y_1\not =0$ and $y_2\not =0$. Let us assume the opposite, that is there is an equilibrium point in $y_1=y_{10}\not =0$ and $y_2=y_{20}\not =0$. Then, $y_1'$ will vanish if $w=\frac{y_{20}}{3 (-1+y_{10}+y_{20})}$. However, introducing this last expression in (\ref{y2p}), it implies that $y_2'$ will vanish only if $y_{20}=0$. This contradicts our assumption that $y_2$ at equilibrium should not be zero. The same thing arises if we determine $w$ from $y_2'$ and try to make vanish $y_1'$ since then $y_{10}$ should be equal to zero at equilibrium. Consequently, the equilibrium points are all such that $y_1=0$ or/and $y_2=0$.\\
Firstly, we consider the case $y_1=0$. Then $y_1'$ disappears if $w$ is such that $y_1(1-y_1-y_2)(1+w)=0$. This is always the case when $w$ is finite. But if $w$ diverges, this equality have to be true otherwise there is no equilibrium point on the line $y_1=0$. $y_2'$ disappears in the following cases, taking into account that $0\leq y_2\leq 1$
\begin{itemize}
\item $y_2=0$ with $w$ such that $y_2(1-y_1-y_2)(1+w)=0$
\item $y_2=y_{20}\not =0$ or $1$ with $w=1/3$
\item $y_2=1$ with $w$ such that $(1-y_1-y_2)(1+w)=0$
\end{itemize}
Secondly, we consider the case $y_2=0$. Then $y_2'$ disappears if $w$ is such that $y_2(1-y_1-y_2)(1+w)=0$. $y_1'$ disappears in the following cases, taking into account that $0\leq y_1\leq 1$
\begin{itemize}
\item $y_1=0$ with $w$ such that $y_1(1-y_1-y_2)(1+w)=0$
\item $y_1=y_{10}\not =0$ or $1$ with $w=0$
\item $y_1=1$ with $w$ such that $(1-y_1-y_2)(1+w)=0$
\end{itemize}
Summarising, we thus have the following equilibrium points
\begin{itemize}
\item $(y_1,y_2)=(1,0)$: this is the usual matter dominated equilibrium point. We call this point $M$. When $w$ diverges, there is a $M$ point if $(1-y_1-y_2)w$ does not diverge.
\item $(y_1,y_2)=(0,1)$: this is the usual radiation dominated equilibrium point. We call this point $R$. When $w$ diverges, there is a $R$ point if $(1-y_1-y_2)w$ does not diverge.
\item $(y_1,y_2)=(0,0)$: this is the dark energy dominated equilibrium point for which Universe expansion can be accelerated (see below). We call this point $D$. When $w$ diverges, there is a $D$ point if $y_1w$ and $y_2w$ do not diverge.
\item $(y_1,y_2)=(y_{10},0)$ with $y_{10}$ a non vanishing constant such that $w(y_{10},0)=0$: at this equilibrium point, dark energy mimics cold dark matter and we call it $DM$. Around $DM$, where $w\simeq 0$, dark energy mimics warm dark matter\cite{Ave12}. In the rest of the paper we will not distinguish between these two cases and we will write that in (when $DM$ is a source or a sink for a trajectory in the phase space) or near (when $DM$ is a saddle for a trajectory) $DM$, dark energy mimics matter and radiation is negligible. If there is no value $y_{10}$ such that $w(y_{10},0)=0$, there is no equilibrium point of this kind. Otherwise there are as many equilibrium points of this kind as there are constant values of $y_1$ making $w(y_1,0)$ vanishing.
\item $(y_1,y_2)=(0,y_{20})$ with $y_{20}$ a non vanishing constant such that $w(0,y_{20})=1/3$: at this equilibrium point, dark energy mimics radiation and we call this kind of point $DR$ (as for $DM$, in the rest of the paper we will not distinguish the cases when $DR$ is a source or a sink for a trajectory with $w=1/3$ or when it is a saddle and $w\simeq 1/3$ during a transient period of time). If there is no value $y_{20}$ such that $w(0,y_{20})=1/3$, there is no equilibrium point of this kind. Otherwise there are as many equilibrium points of this kind as there are constant values of $y_2$ making $w(0,y_2)$ equal to $1/3$.
\end{itemize}
\subsection{General considerations about equilibrium points stability}\label{s12}
\begin{figure}[h]
\centering
\includegraphics[width=6cm]{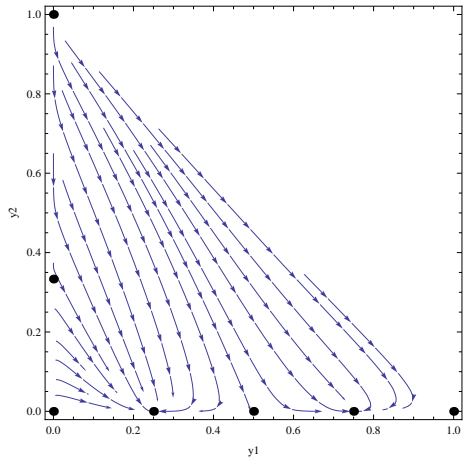}
 \caption{\scriptsize{\label{fig1}Phase plane for $w=(y_1-1/4)(y_1-1/2)(y_1-3/4)\left[(y_2-1/3)-32/9\right]$. There are three equilibrium points in $y_2=0$ such that $w=0$ with $y_1=1/4$, $1/2$ and $1/3$ and one equilibrium point on $y_1=0$ such that $w=1/3$ with $y_2=1/3$.}}
\end{figure}

It is not possible to fully study the stability of the above equilibrium points without specifying $w$. However we can get some general and useful results. 
\begin{itemize}
\item The $M$ point cannot be a source since we assume Universe is expanding and thus at early time, radiation always dominates matter. However, it can be a sink or a saddle. For instance when $w$ and its derivatives $w_{y_1}$ and $w_{y_2}$ do not diverge in $M$, (\ref{sta}) indicates that it is a saddle point when $w(1,0)<0$ and a sink when $w(1,0)>0$.
\item The $R$ point cannot be a sink since we assume Universe is expanding and thus at late time, matter always dominates radiation. However, it can be a source or a saddle. For instance, when $w$ and its derivatives do not diverge in $R$, (\ref{sta}) indicates that it is a saddle point when $w(0,1)>1/3$ and a source when $w(0,1)<1/3$.
\item The $D$ point can be a saddle, a source or a sink. For instance, when $w$ and its derivatives do not diverge in $D$, (\ref{sta}) indicates that it is a saddle point when $0<w(0,0)<1/3$, a source when $w(0,0)>1/3$ and a sink when $w(0,0)<0$.
\item Concerning the $DM$ points, since we assume an expanding Universe, they can never be a source because in $DM$, $y_2/y_1\propto 1+z$ tends to zero. However they can be a sink or a saddle. For instance if $w$'s derivatives do not diverge in $DM$, then from (\ref{sta}) it comes that $DM$ are saddle points when $\frac{dw}{dy_1}(y_{10},0)>0$ and sinks when $\frac{dw}{dy_1}(y_{10},0)<0$.
\item Concerning the $DR$ points, since we assume an expanding Universe , they can never be a sink because in $DR$, $y_2/y_1\propto 1+z$ diverges. However they can be a source or a saddle. Indeed if $w$'s derivatives do not diverge in $DR$, using (\ref{sta}), we deduce that $DR$ are saddle points when $\frac{dw}{dy_2}(0,y_{20})<0$ and sources when $\frac{dw}{dy_2}(0,y_{20})>0$.
\end{itemize}
There is no limit to the number of equilibrium points of the system (\ref{y1p}-\ref{y2p}). For instance, if $w=\prod_{i=1}^n(y_1-y_{1i})\left[\prod_{j=1}^m(y_2-y_{2j})+1/\prod_{i=1}^n (-y_{1i})/3\right]$, there are $R$, $M$ and $D$ equilibrium points but also $n$ equilibrium points $DM_i$ in $(y_1,y_2)=(y_{1i},0)$ with $w=0$, $y_{1i}$ some non vanishing constants and $m$ equilibrium points $DR_j$ in $(y_1,y_2)=(0,y_{2j})$ with $w=1/3$, $y_{2j}$ some non vanishing constants. The case $n=3$ and $m=1$ is plotted on figure \ref{fig1}.\\
From a physical viewpoint, all source (sink) points are such that $y_1=0$ (respectively $y_2=0$). Thus, Universe always starts without matter and ends without radiation, even mimicked by dark energy. Moreover, by definition, Universe expansion accelerates when $d^2a/dt^2=dH/dt+H^2>0$, that is
$$
1/2\left[-3-y_2+3w (-1+y_1+y_2)\right]+1>0
$$
Without surprise this last expression shows that the equilibrium points $M$ and $R$ cannot generate an accelerated expansion whatever the behaviour of $w$ (which cannot diverge faster than $(-1+y_1+y_2)^{-1}$ near $M$ and $R$). $DM$ and $DR$ cannot generate an accelerated expansion neither. This is finally only possible for the equilibrium point $D$ for which $(y_1,y_2)=(0,0)$ and when $w<-1/3$, a well known result \footnote{Other equilibrium points with accelerated expansion could exist with a negative energy density for dark energy but we do not consider this case in that paper}.\\

Last, we examine the dominant energy condition at each equilibrium point. For dark energy, it writes $\mid p \mid/\rho=\mid w \mid \leq 1$. This condition can be violated in $M$, $R$ and $D$, in particular in the case of a ghost dark energy and when $w$ diverges. The condition always holds in $DM$ and $DR$ since then $w$ is $0$ and $1/3$ respectively. If now we consider matter, radiation and dark energy as a unique fluid of density $\rho_{total}=\rho+\rho_m+\rho_r$, then the dominant energy condition writes $\mid p_{total} \mid/\rho_{total}=\mid 1/3y_2+w(1-y_1-y_2)\mid<1$ or $\frac{dH}{dt} =-k/2(p_{total}+\rho_{total})=1/2H^2\left[-3y_1-4y_2-3(1+w)(1-y_1-y_2)\right]<0$ when $p_{total}<0$. These inequalities are always true at the points $M$, $R$, $DM$ and $DR$, even when $w$ diverges in $M$ and $R$ since then equilibrium is reached only if $w<<1/(1-y_1-y_2)$. The dominant energy condition is only violated in $D$, once again when $\mid w\mid>1$. To be more specific about $\frac{dH}{dt}/H^2$, we note that its values in $R$, $DR$, $M$, $DM$ and $D$, are respectively $-2$, $-2$, $-3/2$, $-3/2$ and $-3/2(1+w)$. The acceleration parameter $q=1+\frac{dH}{dt}H^{-2}$ has thus the same negative values in $R$ and $DR$ or in $M$ and $DM$, in agreement with the fact that these two couples of equilibrium points have respectively the dynamics of some radiation and matter dominated epochs.
\section{Universe dynamics}\label{s2}
In this section, we look for some sequences of epochs describing a Universe starting by some early radiation dominated epochs, followed by some matter dominated epochs and ending by an accelerated expansion. Such sequences are described in the phase space $(y_1,y_2)$ by trajectories having a source in $R$ or $DR$, passing near some saddle points $R$ or $DR$, then near some saddle points $M$ or $DM$ and ending in $D$. For instance, a phase space trajectory starting in $R$, repelled by a first saddle point $DR$, then a second saddle point $DM$ and ending in $D$ will be noted $R\rightarrow DR \rightarrow  DM \rightarrow D$. There are an infinite number of possibilities depending on the form of $w$. However we will show in the next section that if $w$ is not singular at some saddle equilibrium points, there are only ten of such sequences. We present them in this section by using polynomial forms of $w(y_1,y_2)$. They are such that $w$ and its derivatives stay finite whatever $0\leq y_i\leq 1$. This also allows to use results of section \ref{s12} and to get easily some $w$ reproducing the above desired sequences of epochs.
\subsection{Three successive epochs}\label{s21}
\begin{figure}[h!]
\centering
\includegraphics[width=6cm]{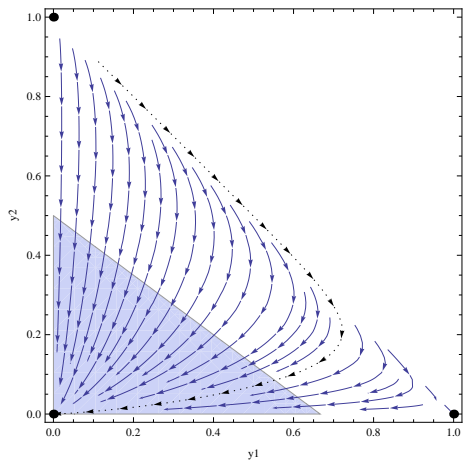}
\caption{\scriptsize{\label{fig3}Phase plane for the cosmological constant $w=-1$. Grey area is the zone of the phase plane where Universe expansion is accelerated. The points are the equilibrium points. A trajectory is plotted as a dotted line.}}
\end{figure}

\begin{figure}[h!]
\centering
\includegraphics[width=6cm]{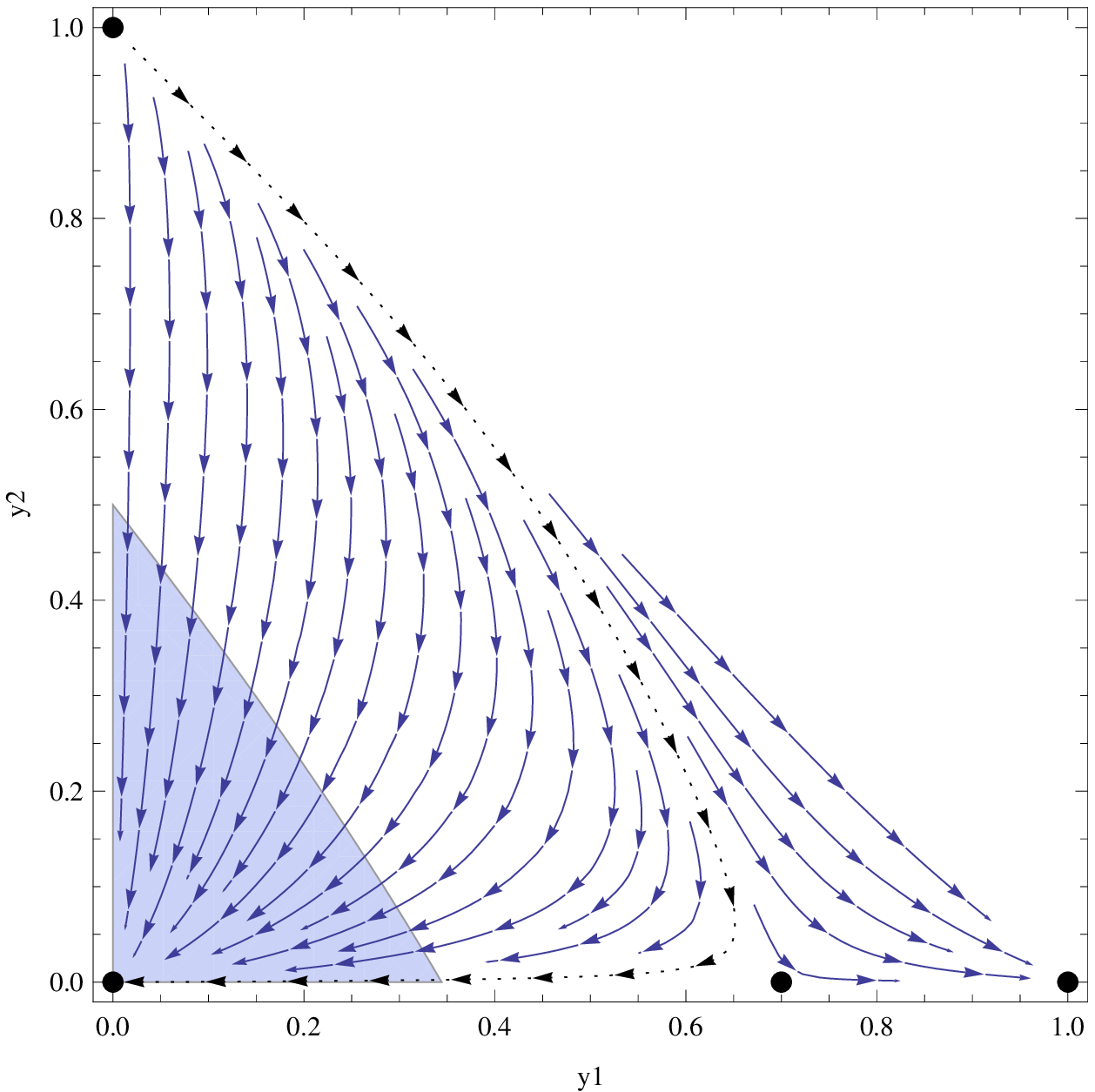}
\caption{\scriptsize{\label{fig4}Phase plane for $w=-1+1.42 y_1$ with a saddle equilibrium point $DM$ in $(y_1,y_2)=(0.7,0)$. Grey area is the zone of the phase plane where Universe expansion is accelerated. The points are the equilibrium points. A trajectory is plotted as a dotted line.}}
\end{figure}

\begin{figure}[h!]
\centering
\includegraphics[width=6cm]{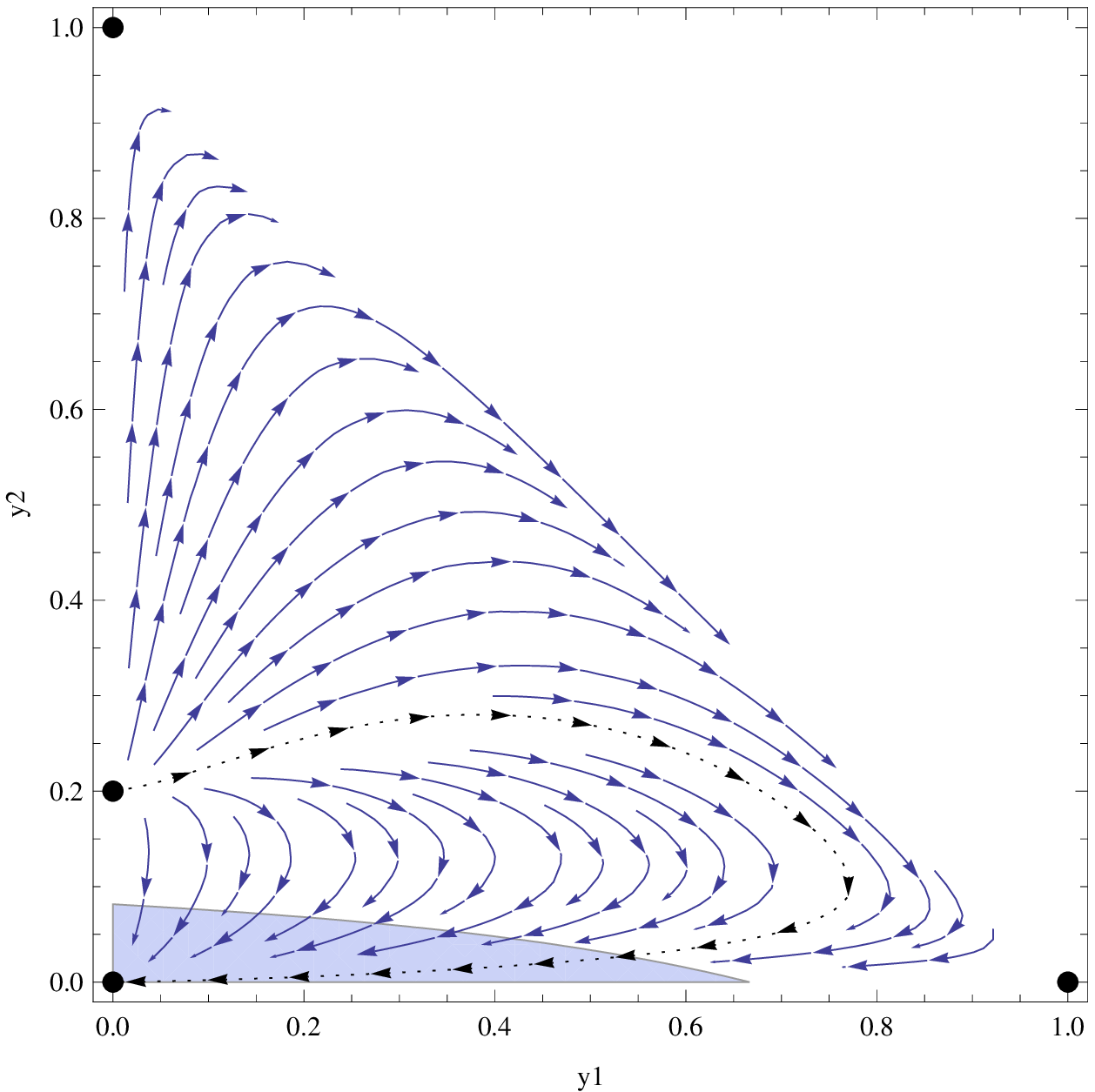}
\caption{\scriptsize{\label{fig5}Phase plane for $w=-1+7.98 y_2 -6.58 y_2^2$ with a source equilibrium point $DR$ in $(y_1,y_2)=(0,0.2)$. Grey area is the zone of the phase plane where Universe expansion is accelerated. The points are the equilibrium points. A trajectory is plotted as a dotted line.}}
\end{figure}

\begin{figure}[h!]
\centering
\includegraphics[width=6cm]{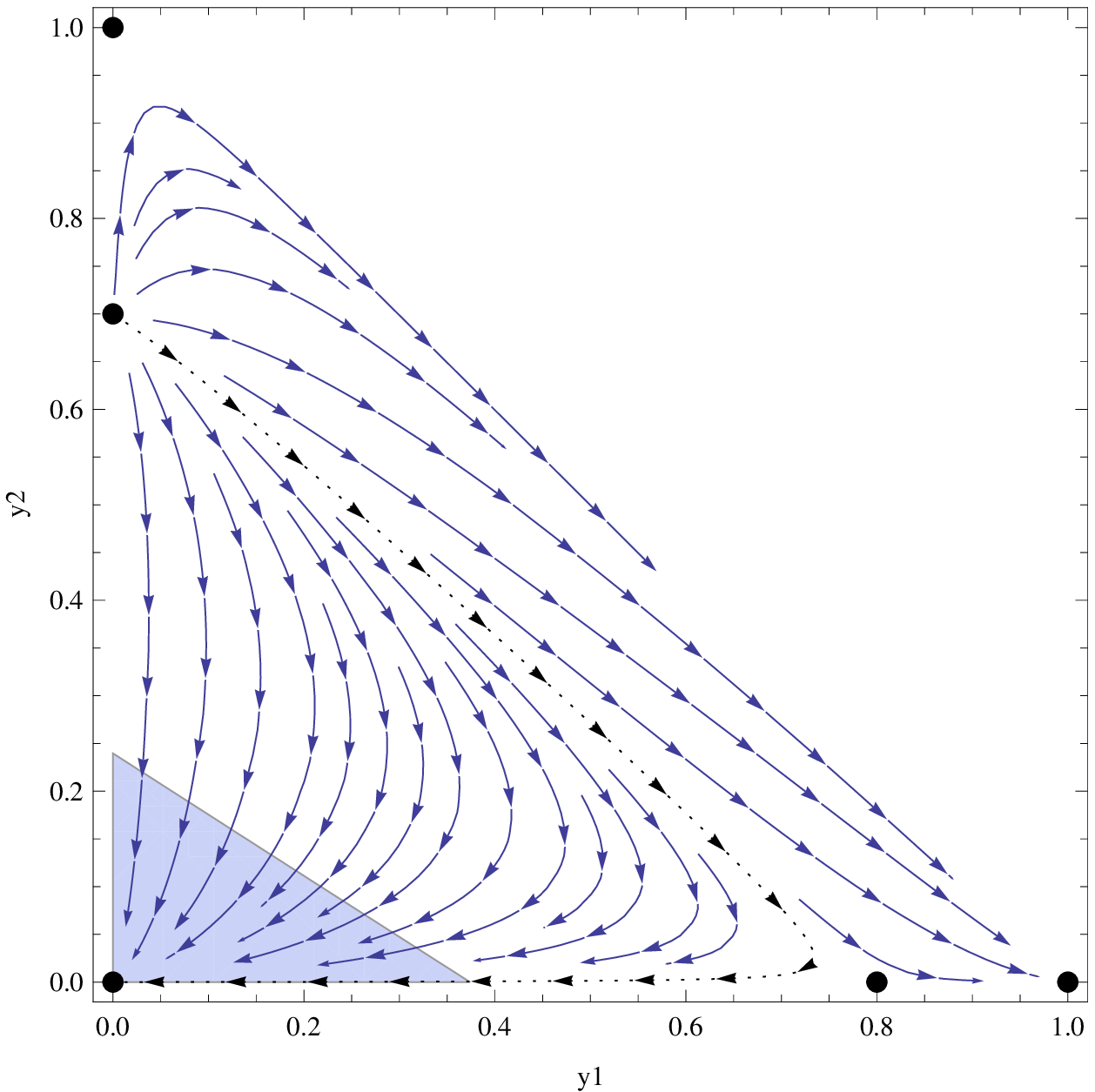}
\caption{\scriptsize{\label{fig6}Phase plane for $w=-1+1.25 y_1 +1.90 y_2$ with a saddle equilibrium point $DM$ in $(y_1,y_2)=(0.8,0)$ and a source equilibrium point $DR$ in $(y_1,y_2)=(0,0.7)$. Grey area is the zone of the phase plane where Universe expansion is accelerated. The points are the equilibrium points. A trajectory is plotted as a dotted line.}}
\end{figure}

In this subsection, we consider the above described sequences with three successive epochs. Without surprise, there are four possibilities that we name $S_{31}$ to $S_{34}$ (where in $S_{ij}$, $i$ stands for the number of successive epochs in a sequence, and $j$ is the sequence number):\\\\
$\bullet\mbox{ $S_{31}$: } R \rightarrow  M \rightarrow   D$\\
This is the sequence followed by the $\Lambda CDM$ model illustrated on figure \ref{fig3} and most of the dark energy models described in the literature. 
\\\\
$\bullet\mbox{ $S_{32}$: } R \rightarrow  DM \rightarrow   D$\\
Dark energy mimics dark matter at intermediate time and accelerates expansion at late time. There is thus at least one value $y_{10}$ of $y_1$ such that $w(y_{10},0)=0$. Although one could think that something similar occurs with the Chaplygin gas\cite{Kam01}, this last theory rather follows the $S_{31}$ sequence with $w=0$ in $(y_1,y_2)=(0,1)$, i.e. at early times. An example of sequence $S_{32}$ is shown on figure \ref{fig4}.
\\\\
$\bullet\mbox{ $S_{33}$: } DR \rightarrow  M \rightarrow   D$\\
Dark energy mimics radiation at early time and accelerates expansion at late time. There is thus at least one value $y_{20}$ of $y_2$ such that $w(0,y_{20})=1/3$. This kind of dark energy plays the same role as Chaplygin gas at early time but mimicking radiation ($w\rightarrow 1/3$) instead of matter. An example of such a sequence is shown on figure \ref{fig5}.
\\\\
$\bullet\mbox{ $S_{34}$: } DR \rightarrow  DM \rightarrow   D$\\
Universe evolution is driven by dark energy that mimics radiation at early time, (dark) matter at intermediate time and accelerates expansion at late time. In this case, there are at least one value $y_{10}$ of $y_1$ and one value $y_{20}$ of $y_2$ such that respectively $w(y_{10},0)=0$ and $w(0,y_{20})=1/3$. An example of such a sequence is shown on figure \ref{fig6}.
\\\\
To our knowledge, cosmological models following sequences $S_{32}$ to $S_{34}$ have not been studied in the literature. Indeed, most of dark energy models focused on late time epoch. Consequently the role plays by a dark energy as an additional source of radiation at early times or/and of dark matter at intermediate times, is usually not envisaged.
\subsection{Four successive epochs}\label{s22}
\begin{figure}[h!]
\centering
\includegraphics[width=6cm]{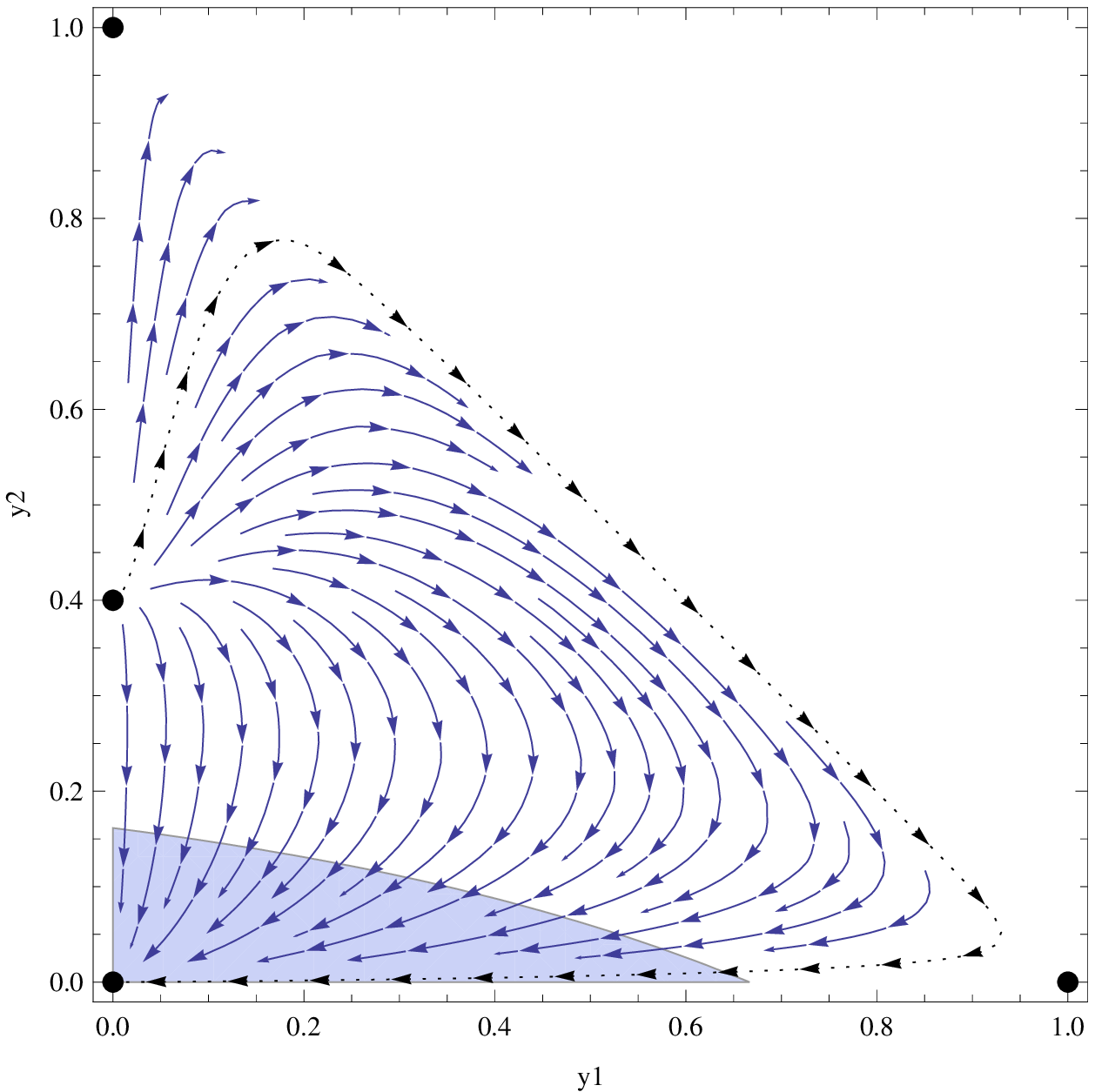}
\caption{\scriptsize{\label{fig7}Phase plane for $w=-1+3.33 y_2$ with a source equilibrium point $DR$ in $(y_1,y_2)=(0,0.4)$. Grey area is the zone of the phase plane where Universe expansion is accelerated. The points are the equilibrium points. A trajectory is plotted as a dotted line.}}
\end{figure}

\begin{figure}[h!]
\centering
\includegraphics[width=6cm]{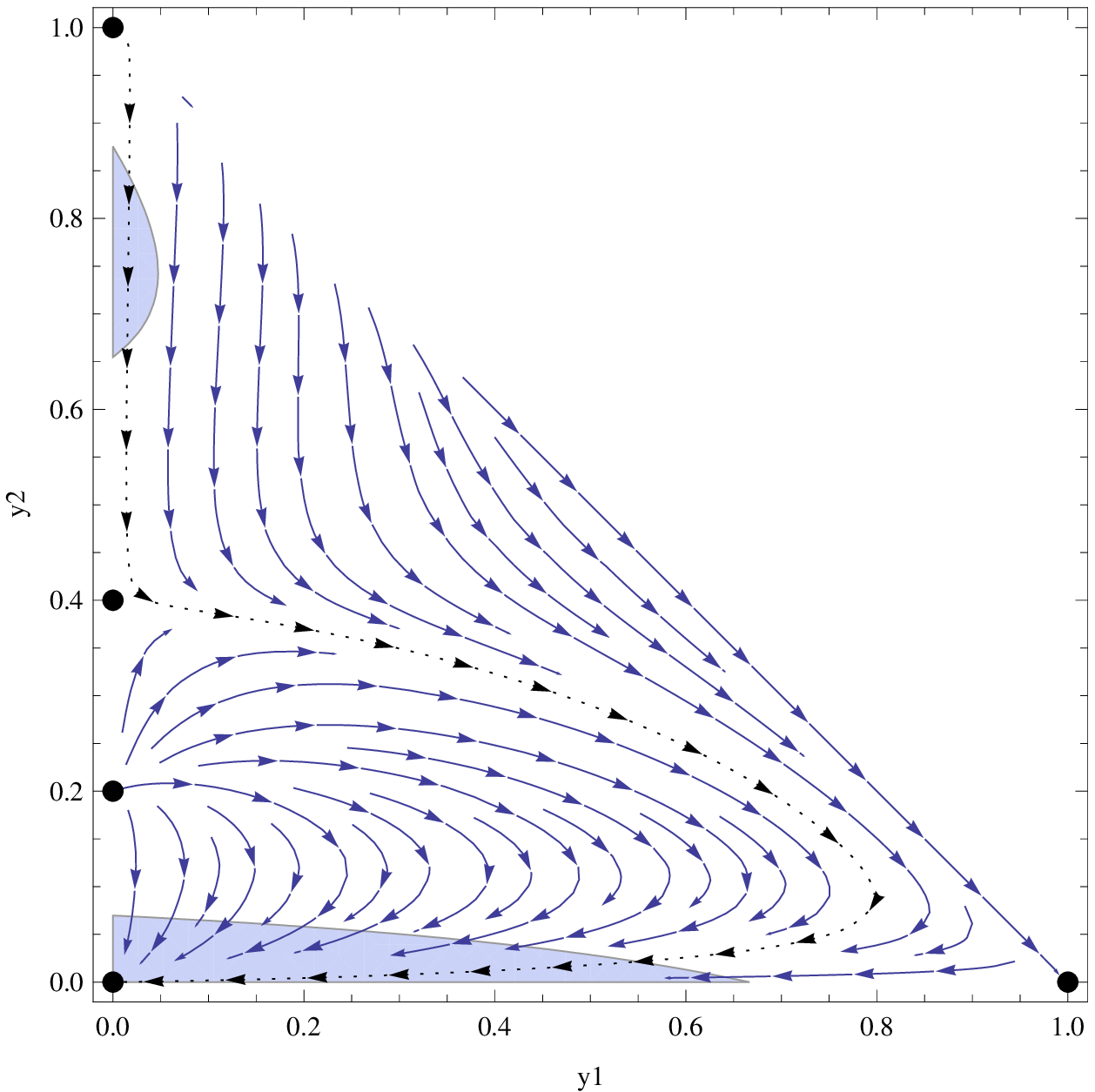}
\includegraphics[width=6cm]{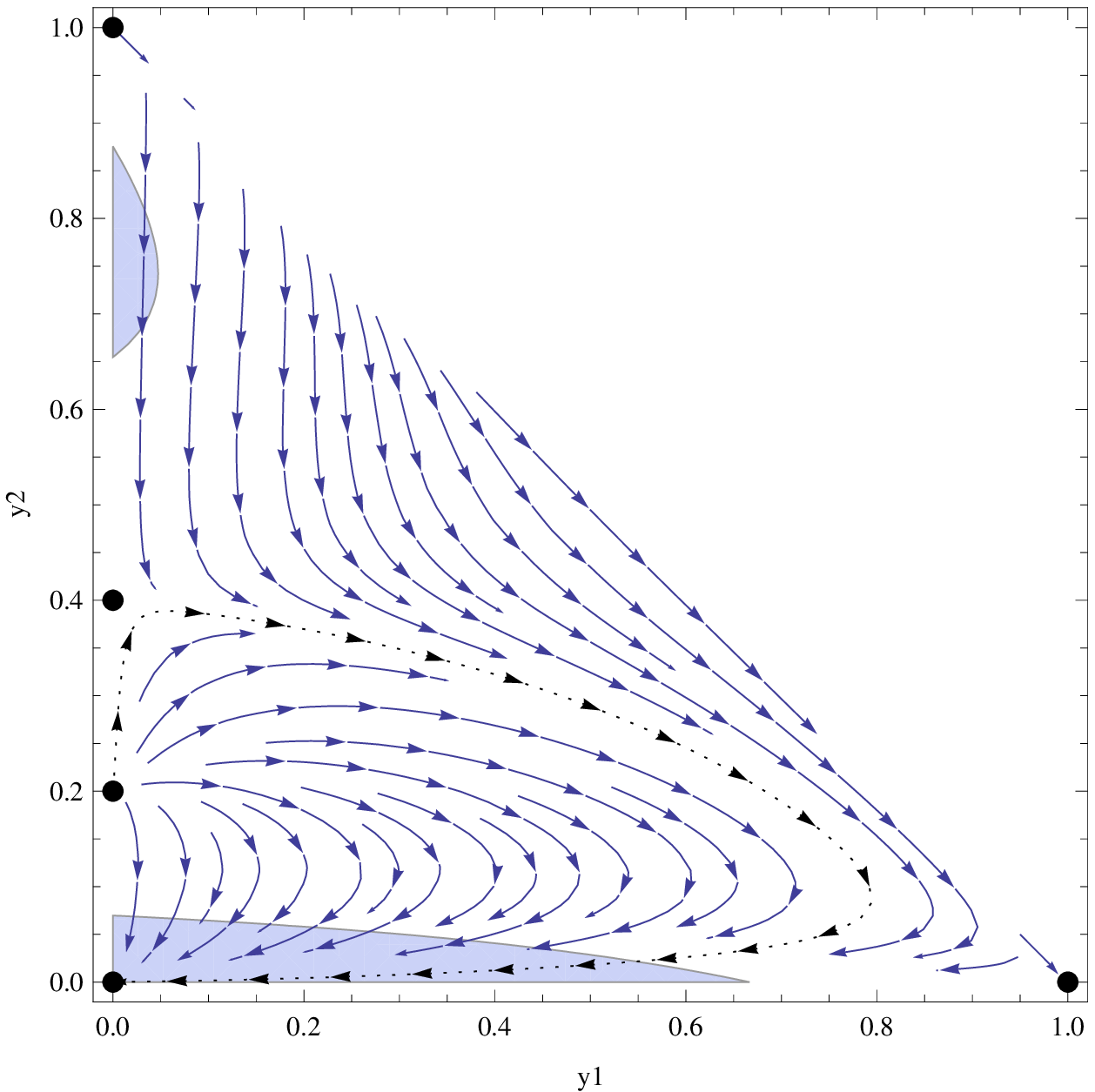}
\caption{\scriptsize{\label{fig8}Phase plane for $w=-1+10 y_2-16.66 y_2^2$ with two equilibrium points $DR$, one saddle in $(y_1,y_2)=(0,0.4)$ and one source in $(y_1,y_2)=(0,0.2)$. Grey area is the zone of the phase plane where Universe expansion is accelerated. The points are the equilibrium points. Two trajectories are plotted as dotted lines. The first panel shows a $R \rightarrow DR \rightarrow  M \rightarrow   D$ trajectory and the second panel a $DR \rightarrow DR \rightarrow  M \rightarrow   D$ trajectory.}}
\end{figure}

\begin{figure}[h!]
\centering
\includegraphics[width=6cm]{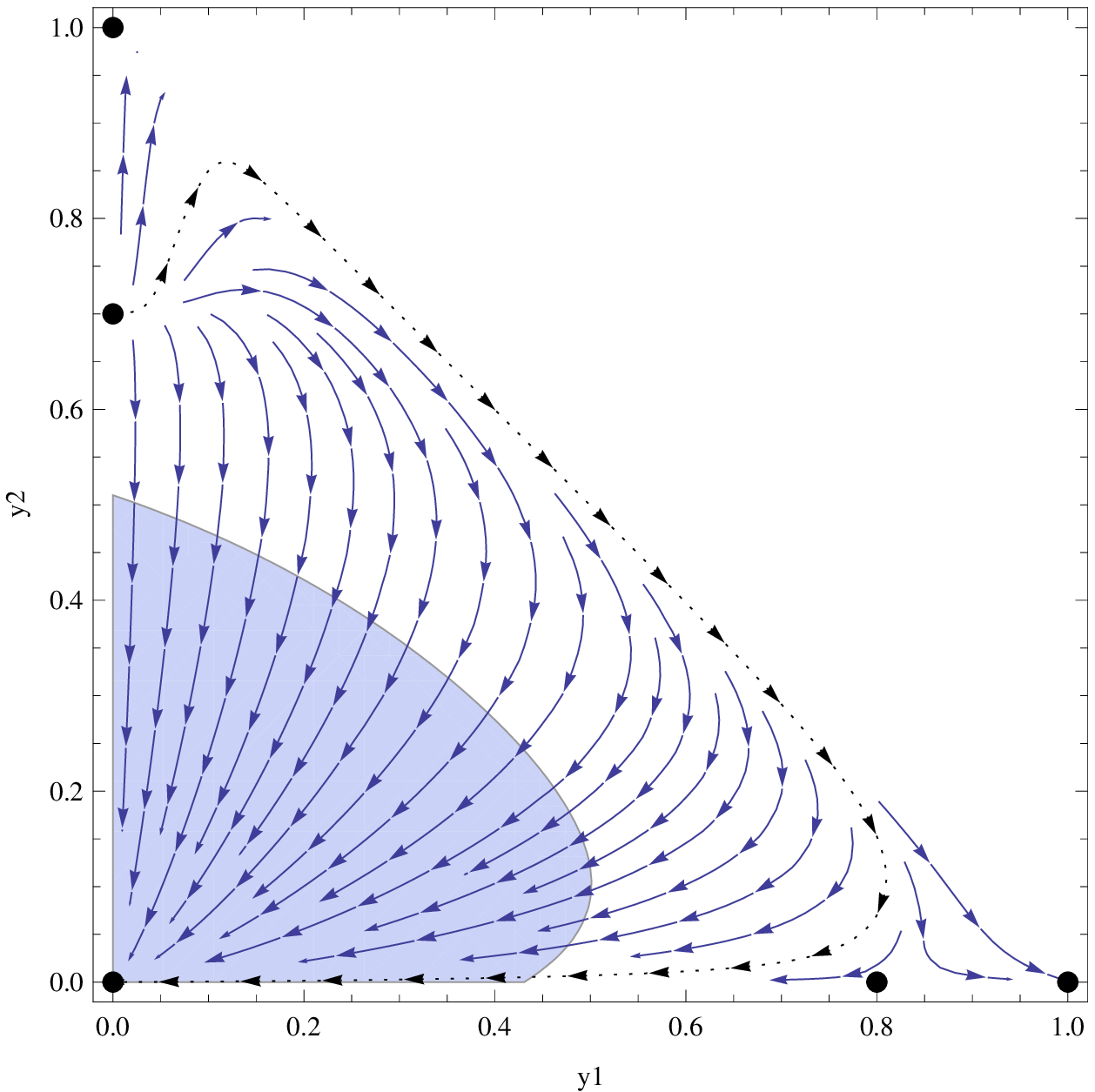}
\caption{\scriptsize{\label{fig10}Phase plane for $w=-1+0.62 y_1+0.78 y_1^2-5.31 y_2+10.31 y_2^2$ with a saddle equilibrium points $DM$ in $(y_1,y_2)=(0.8,0)$ and a source equilibrium point $DR$ in $(y_1,y_2)=(0,0.7)$. Grey area is the zone of the phase plane where Universe expansion is accelerated. The points are the equilibrium points. A trajectory is plotted as a dotted line.}}
\end{figure}

\begin{figure}[h!]
\centering
\includegraphics[width=6cm]{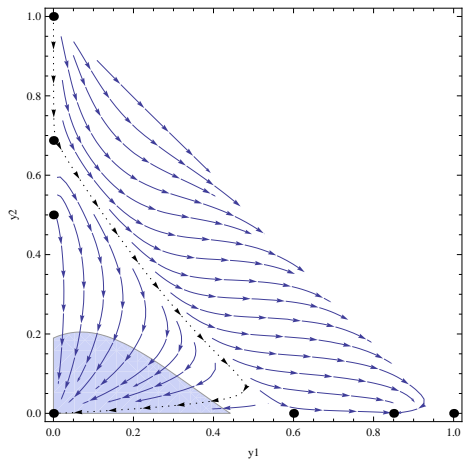}
\caption{\scriptsize{\label{fig11}Phase plane for $w=-3/2-3y_1+17.66 y_1^2-14.16 y_1^3+6.33 y_2-5.33 y_2^2$ with two equilibrium points $DM$, one sink in $(y_1,y_2)=(0.85,0)$ and one saddle in $(y_1,y_2)=(0.6,0)$, and two equilibrium points $DR$, one source in $(y_1,y_2)=(0,0.5)$ and one saddle in $(y_1,y_2)=(0,0.68)$. Grey area is the zone of the phase plane where Universe expansion is accelerated. The points are the equilibrium points. A trajectory is plotted as a dotted line.}}
\end{figure}

\begin{figure}[h!]
\centering
\includegraphics[width=6cm]{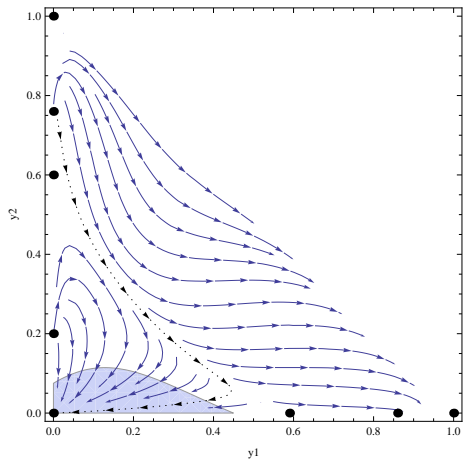}
\caption{\scriptsize{\label{fig12}Phase plane for $w=-1.12-5.74y_1+23.12y_1^2-17.29y_1^3+11.61y_2-24.70 y_2^2+15.72 y_2^3$ with two equilibrium points $DM$, one sink in $(y_1,y_2)=(0.86,0)$ and one saddle in $(y_1,y_2)=(0.59,0)$, and three equilibrium points $DR$, two sources in $(y_1,y_2)=(0,0.76)$ and $(y_1,y_2)=(0,0.2)$ and one saddle in $(y_1,y_2)=(0,0.6)$. Grey area is the zone of the phase plane where Universe expansion is accelerated. The points are the equilibrium points. A trajectory is plotted as a dotted line.}}
\end{figure}

In this subsection, we consider sequences with four successive epochs. There should be $12$ possibilities but only $6$ of them exist when $w$ is not singular at some saddle equilibrium points (see section \ref{s3}).
\\\\
$\bullet\mbox{ $S_{41}$: } DR \rightarrow R \rightarrow  M \rightarrow   D$\\
This first possibility is similar to usual dark energy models but there are now two radiation epochs at early times. For the first one, there must be a value $y_{20}$ of $y_2$ such that $w(0,y_{20})=1/3$ and then, dark energy mimics radiation. An example of such a sequence is plotted on figure \ref{fig7}.
\\\\
$\bullet\mbox{ $S_{42}$: } R \rightarrow DR \rightarrow  M \rightarrow   D$\\
This sequence is similar to $S_{41}$ but the two radiation epochs are reversed. An example of such a sequence is plotted on the first panel of figure \ref{fig8}.
\\\\
$\bullet\mbox{ $S_{43}$: } DR \rightarrow DR \rightarrow  M \rightarrow   D$\\
During the two radiation epochs dark energy mimics radiation but in different proportions. This time we thus need two non vanishing values of $y_2$ such that $w(0,y_2)=1/3$. An example of such a sequence is plotted on the second panel of figure \ref{fig8}.\\\\
$\bullet\mbox{ $S_{44}$: } DR \rightarrow R \rightarrow  DM \rightarrow   D$\\
Dark energy mimics radiation during the first radiation epoch. The second epoch is dominated by pure radiation. During the third epoch, dark energy mimics matter. Thus there should be at least one value $y_{10}$ of $y_1$ and one value $y_{20}$ of $y_2$ such that respectively $w(y_{10},0)=0$ and $w(0,y_{20})=1/3$. An example of such a sequence is plotted on figure \ref{fig10}.
\\\\
$\bullet\mbox{ $S_{45}$: } R \rightarrow DR \rightarrow  DM \rightarrow   D$\\
This is similar to $S_{44}$ but with the two radiation epochs reversed. An example of such a sequence is plotted on figure \ref{fig11}.
\\\\
$\bullet\mbox{ $S_{46}$: } DR \rightarrow DR \rightarrow  DM \rightarrow   D$\\
Dark energy mimics radiation during the first and second epoch, matter during the third epoch and finally accelerates expansion at late time. All Universe dynamics is thus driven by dark energy. At least two values of $y_2$ have to be such that $w(0,y_2)=1/3$ and one value of $y_1$ such that $w(y_1,0)=0$. An example of such a sequence is plotted on figure \ref{fig12}.
\\\\
In the next section, we show that these $10$ sequences, that describe a Universe starting with some radiation epochs followed by some matter epochs and ending by an accelerated expansion, are the only ones that can be got if $w$ is not singular at some saddle equilibrium points.
\section{Discussion}\label{s3}
In the previous section, we reproduced $10$ sequences with no more than four epochs describing Universe starting with some radiation epoch(s) followed by some matter epoch(s) and ending by an accelerated expansion. But, of course, there are an infinity of them, with as many epochs as desired. Indeed one can always choose any form of the scale factor $a$ and then derive the corresponding $w$. We remark that a common property of this infinity of sequences, but the ten we found in the previous section, is that they always contain two successive saddle epochs dominated by matter (noted as $DM \leftrightarrow  M$ or $DM \leftrightarrow DM$) or radiation (noted as $DR \leftrightarrow  R$ or $DR \leftrightarrow  DR$). In the same time we show in the appendix that when $w$ and/or its derivatives with respect to $y_1$ and $y_2$ are finite, it is not possible to get two successive saddle epochs dominated by matter or radiation. Consequently, we deduce that to get additional sequences of epochs with respect to the ten we found in section \ref{s2}, $w$ and/or its first derivatives with respect to $y_1$ and/or $y_2$ have to diverge at some saddle equilibrium points of type $M$, $DM$, $R$ or $DR$ and thus be singular at some of these points. From a physical viewpoint, since a trajectory pass near a saddle point and not exactly through it, it means that the $w$ corresponding to this trajectory will have some pronounced bumps or dips.\\
\\
Such a behaviour seems not favoured by observations. Hence in \cite{Bie13}, no evidence is found for a spike in the equation of state at early time that would have left an imprint on the CMB anisotropy pattern and the rate of growth of large scale structure. In \cite{Mor09} it is shown that large fluctuations of the equation of state are allowed only at ultra-low redshifts $z<0.02$. Another way to interpret the divergence of $w$ and its derivatives is to consider the adiabatic sound speed. It writes
$$
c_s^2=\frac{dp/dt}{d\rho/dt}=w-\frac{w'}{3(1+w)}
$$
or with the normalised variables
\begin{equation}\label{cs2}
c_s^2=w+\frac{y_2 (w_{y_2}-w_{y_1}y_1-w_{y_2}y_2)+3 w (-1+y_1+y_2) (w_{y_1}y_1+w_{y_2}y_2)}{3 (1+w)}
\end{equation}
When $c_s^2>1$, perturbations move at superluminal speeds\cite{Bon06} whereas when $c_s^2<0$, the system is perturbatively unstable on small scales and therefore unphysical. Then, as explained in \cite{PerBac02}, the adiabatic pressure would accelerate the collapse of density perturbations unless entropic perturbations be taken into account and the the sound speed of perturbations be no more the adiabatic sound speed. Also, we are going to examine the values of $c_s^2$ at each equilibrium point to check when $0<c_s^2<1$.\\
Firstly, we assume that $w$ and its derivatives $w_{y_i}$ do not diverge. Then $c_s^2=w$ for all the equilibrium points. It will be negative in $M$ and $D$ when the first point is a saddle and the second one a sink since then $w<0$ (see subsection \ref{s12}). In $R$, there is no restriction on the sign of $w$ when this point is a source or a saddle. However, we may have $c_s^2<0$ only when $R$ is a source or $c_s^2>1$ only when $R$ is a saddle. There is no problem for $c_s^2$ in $DM$ and $DR$ since $w=0$ and $w=1/3$ respectively. \\
Secondly, we assume that $w$ does not diverge but its derivatives $w_{y_i}$. Then $\mid c_s^2\mid >>1$ in $M$ when the $w_{y_i}$ diverge faster than $1/y_2$, in $DM$ when $w_{y_i}>>1/y_2$ and $w_{y_1}>>1/w$, in $R$ or $DR$ when the $w_{y_i}$ diverge faster than $1/y_1$ or finally in $D$ when $w_{y_i}$ diverges faster than $1/y_i$ with $i=1,2$. If these inequalities are not respected, then the divergence of the $w_{y_i}$ do not intervene in the value of $\mid c_s^2\mid$ and we recover the previous results, when the $w_{y_i}$ do not diverge.\\
Last, when $w$ diverges at equilibrium, generally $c_s^2$ always diverges (but in the special case for which the divergence of $w$ is compensated by this of the second term in the right side member of equation (\ref{cs2})).\\ 
Summarising, if we assume that $M$ is a saddle and $D$ is a sink, $c_s^2$ is always negative at these points, a well known fact for $D$, underlined in \cite{PerBac02}. For the other equilibrium points, we will have $0<c_s^2<1$ only if $w$ does not diverge and then in $R$ if $w_{y_i}<<1/y_1$ and $0<w<1$, in $DM$ if $w_{y_i}<<1/y_2$ and $w_{y_1}<<1/w$, in $DR$ if $w_{y_i}<<1/y_1$. Hence divergence of $w$ and/or its derivatives $w_{y_i}$ generally reduces the number of equilibrium points where the adiabatic sound speed is well defined whereas we will always have to take into account entropic perturbations and to consider an effective sound speed\cite{PerBac02} at intermediate and late times in $M$ and $D$ respectively, to study structures formation in presence of dark energy.\\

\begin{figure}[h!]
\centering
\includegraphics[width=6cm]{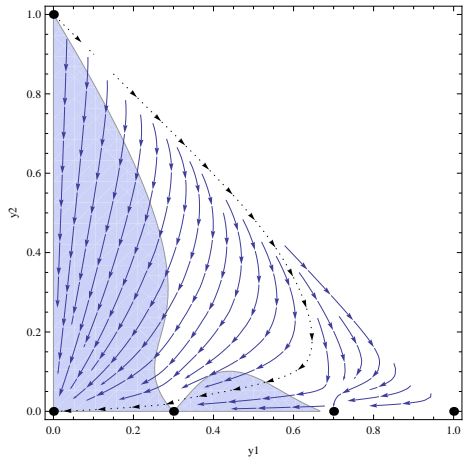}
\caption{\scriptsize{\label{fig14}Phase plane for $w=-(y_1-0.3)^2(y_1-0.7)^2/((y_1-0.3)^2 + y_2^2)/((y_1-0.7)^2+y_2^2) - y_2/y_1$ with two $DM$ equilibrium points in $(y_1,y_2)=(0.3,0)$ and $(y_1,y_2)=(0.7,0)$. The first one is a saddle. The second one is a saddle when $y_1<0.7$ and a sink otherwise. Grey area is the zone of the phase plane where Universe expansion is accelerated. The points are the equilibrium points. A trajectory presenting a sequence $R\rightarrow DM\rightarrow DM\rightarrow D$ is plotted as a dotted line.}}
\end{figure}

\begin{figure}[h!]
\centering
\includegraphics[width=6cm]{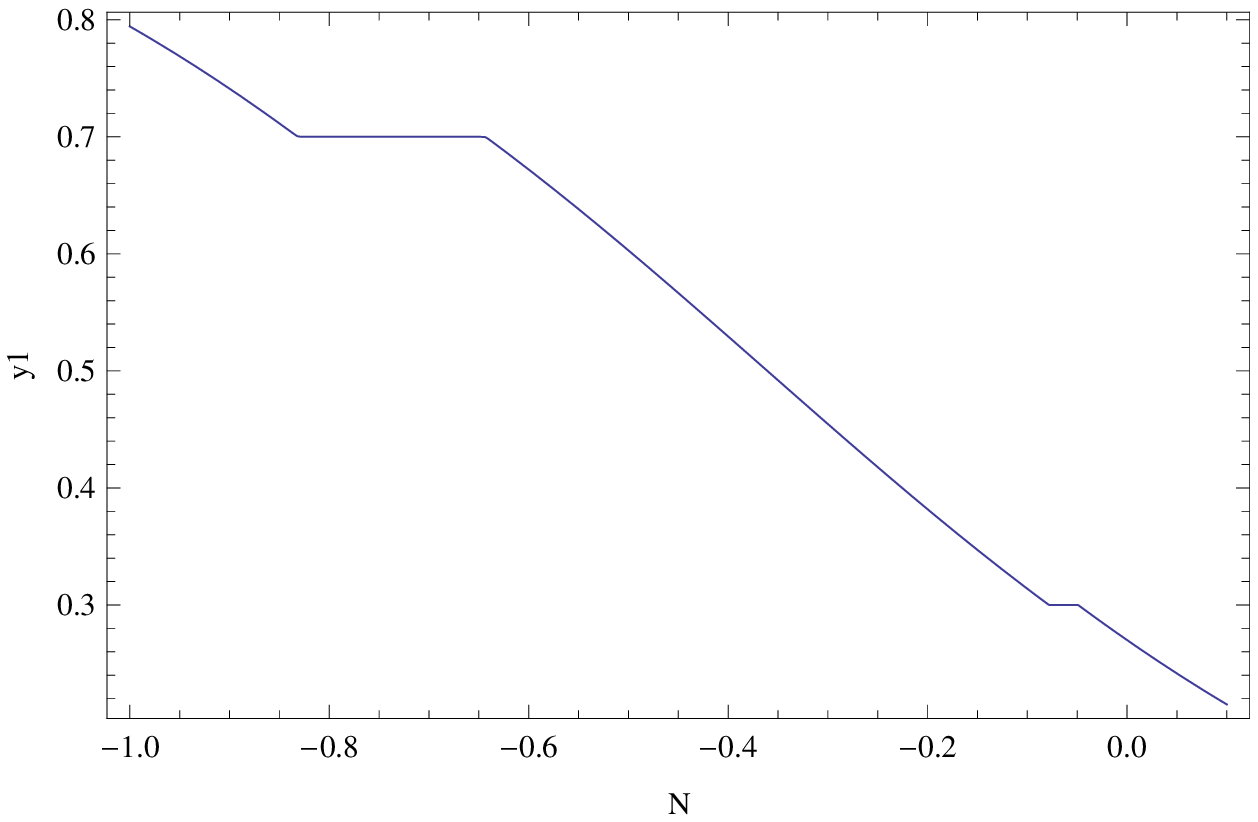}
\includegraphics[width=6cm]{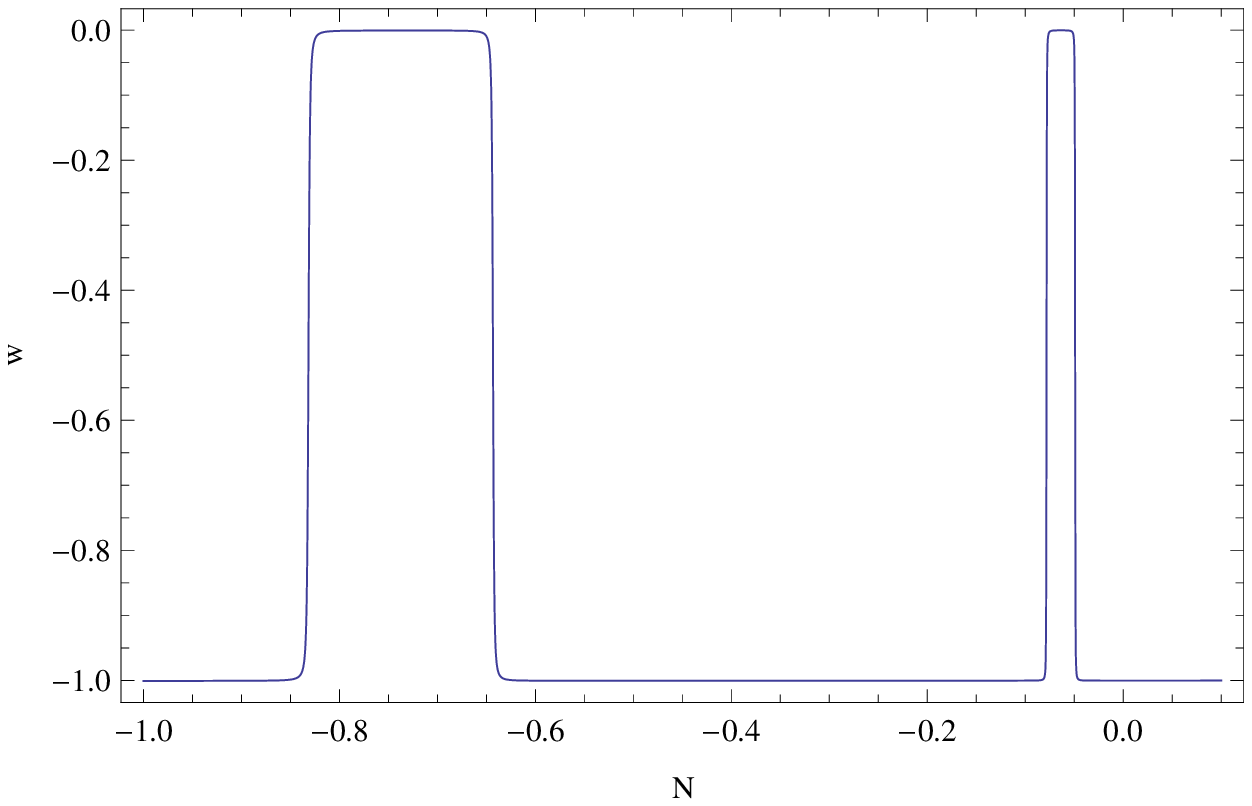}
\caption{\scriptsize{\label{fig15} $y_1(N)$ and $w(N)$ when $w=-(y_1-0.3)^2(y_1-0.7)^2/((y_1-0.3)^2 + y_2^2)/((y_1-0.7)^2+y_2^2) - y_2/y_1$ with $y_1(0)=0.27$ and $y_2(0)=8.27\times 10^{-5}$. The two bumps in the evolution of $w$ correspond to the trajectory in the phase space approaching the $DM$ saddle points.}}
\end{figure}

To illustrate the fact that a form of $w$ with singular behaviours at some equilibrium points $M$, $R$, $DM$ and $DR$ allows to get some sequences different from the ten we present in the previous sections, we try many forms of $w$ with such properties. This is generally not enough to get these new sequences. This implies that a singular $w$ at some equilibrium points is a necessary but not sufficient conditions to get them. Finally, we found that the function $w$ given by
$$
w=-(y_1-0.3)^2(y_1-0.7)^2/((y_1-0.3)^2 + y_2^2)/((y_1-0.7)^2+y_2^2) - y_2/y_1
$$
allows to reproduce a new sequence, the sequence $R\rightarrow DM\rightarrow DM\rightarrow D$. $w$ is zero in $(y_1,y_2)=(0.3,0)$ and $(y_1,y_2)=(0.7,0)$ but its derivatives are diverging at these points (see below). The corresponding phase space is plotted on figure \ref{fig14}.
A sequence $R\rightarrow DM\rightarrow DM\rightarrow D$ is shown as a dotted line but is difficult to observe clearly. Also we plotted on figure \ref{fig15} the evolution of $y_1(N)$ and $w(N)$ when $y_1(0)=0.27$ and $y_2(0)=8.27\times 10^{-5}$, the present day observational values for matter and radiation density parameters. $w$ has two pronounced (but not singular since the trajectory pass close but not exactly on the $DM$ points) bumps as shown by the second graph on figure \ref{fig15}. These bumps correspond to the trajectory approaching the two $DM$ saddle points in $(y_1,y_2)=(0.3,0)$ and $(y_1,y_2)=(0.7,0)$, where $w$ is zero and its derivatives are diverging. If these bumps are so pronounced, this is due to the fact that radiation density parameter should be very small during matter era. Hence the trajectory pass very close to the points where $w$ is singular. Note that between the saddle points, the expansion undergoes a transient acceleration with $w=-1$ that does not correspond to an equilibrium state.
\\
\\
To conclude, in this paper we looked for sequences of epochs able to reproduce our Universe history when gravity is described by General Relativity with radiation, matter and a dark energy defined by an equation of state $w$. We assumed that Universe is expanding and dark energy density is positive. Using dynamical systems methods and a set of normalised variables, we defined all the possible equilibrium points of this theory with these variables and gave some general properties about their stability. These equilibrium points describe some classical matter or radiation dominated epochs but also dark energy epochs during which dark energy mimics matter, radiation or accelerates Universe expansion. When $w$ is not singular at some saddle equilibrium points, there are ten possible sequences of epochs describing our Universe history. There are four sequences with three epochs and six sequences with four epochs. In the dark energy literature, cosmological models often follow sequence $S_{31}$. Here we find some new possibilities for which dark energy can also mimic radiation in a similar way it mimics matter in the Chaplygin model. Although these possibilities may not be distinguished by observations like supernovae, they could be of interest from structure formation viewpoint.\\
To get more than these ten sequences, it is necessary but not sufficient that $w$ or/and its first derivatives with respect to $y_1$ and/or $y_2$  diverge at some equilibrium points $M$, $DM$, $R$ or/and $DR$, i.e. at intermediate time. However such properties for $w$ seem to be discarded by observations\cite{Bie13}\cite{Mor09}. These properties also increase the number of equilibrium points where the adiabatic sound speed is ill-defined. Moreover, the divergence of $w$ leads to the violation of the dominant energy condition. All these clearly distinguish the ten above sequences from an infinity of ways to describe Universe expansion.
\appendix
\section*{Appendix - Special case: $w(y_1,y_2)$ and its derivatives are finite}
\begin{figure}[h]
\centering
\includegraphics[width=6cm]{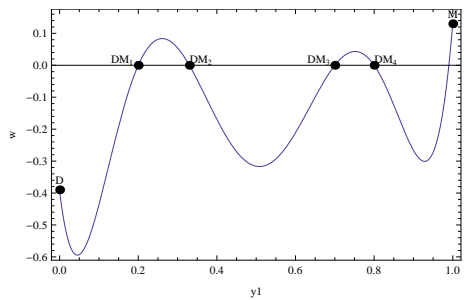}
 \caption{\scriptsize{\label{fig13}The intersection of a surface $w(y_1,y_2)$ with the plane $y_2=0$ forms a curve $(C_M)$ containing all the equilibrium points with $y_2=0$.}}
\end{figure}

Here we show that when $w$ and its derivatives are finite, we cannot have two successive saddle epochs dominated by matter or radiation, i.e we cannot have a transition between equilibrium points like $DM \leftrightarrow  DM$, $DR \leftrightarrow  DR$, $DM \leftrightarrow  M$ or $DR \leftrightarrow  R$. For that we use the results of section \ref{s12} where the stability of the equilibrium points is given in such a case.\\
Let us consider the equation of state $w(y_1,y_2)$. $w$ and its derivatives with respect to $y_1$ and $y_2$ are finite everywhere on the phase plane. We define the intersection of the surface $w(y_1,y_2)$ with the plane $y_2=0$ as a $C^1$ curve $(C_M)$ containing all the equilibrium points with $y_2=0$, i.e. the points $D$, $M$ and $n$ points of type $DM$. We call these last successive points $DM_1$, $DM_2$...$DM_n$. $DM_1$ is the nearest equilibrium point from $D$ and $DM_n$ the nearest equilibrium point from $M$. This is illustrated on figure \ref{fig13} with $n=4$. In the same way, we define $(C_R)$ the intersection of the surface $w(y_1,y_2)$ with the plane $y_1=0$. It thus contains all the equilibrium points $D$, $R$, $DR_1$...$DR_m$ with $y_1=0$.\\
First, we show that when $DM_i$ is a saddle then $DM_{i+1}$ is a sink and $DM_{i+2}$ is a saddle again. When $DM_i$ is a saddle, $w=0$ and $\frac{dw}{dy_1}(y_{10},0)>0$ at this point. Following the $(C_M)$ curve, it will necessarily cross the $w=0$ value again in $DM_{i+1}$ with this time $\frac{dw}{dy_1}(y_{10},0)<0$ like on figure \ref{fig13}. Hence $DM_{i+1}$ will be a sink. The same reasoning leads to show that in $DM_{i+2}$, $\frac{dw}{dy_1}(y_{10},0)>0$ and it will be a saddle. Hence we have shown that when there are several $DM$ points on $y_2=0$, if the first one is a saddle (sink), the next one is a sink (respectively saddle) and so on. In the same way one can prove that if there are several $DR$ points on $y_1=0$, if the first one is a saddle (source), the next one is a source (respectively saddle) and so on. It follows that when $w$ and its derivatives are finite everywhere, we cannot have two successive saddle epochs like $DM_i \rightarrow DM_j$ because any trajectory going from one saddle point $DM_i$ to the other saddle point $DM_j$ would isolate a sink point between $DM_i$ and $DM_j$ from any source. One shows in the same way that the transition between two successive saddle epochs like $DR_i \rightarrow DR_j$ is not possible because any trajectory going from one saddle point to the other would isolate a source point between $DR_i$ and $DR_j$ from any sink.\\
Now we want to show that if $DM_n$ is a saddle point, the next equilibrium point on a trajectory cannot be a saddle point of type $M$ and vice-versa. Indeed if the point $DM_n$ is saddle, then in $DM_n$, $w=0$ and $\frac{dw}{dy_1}(y_{10},0)>0$. Following the continuous curve $(C_M)$, we thus arrive in $M$ with $w>0$ \footnote{Otherwise, $w$ would be equal to zero between $DM_n$ and $M$ on $(C_M)$ and there will be another $DM$ point closest to $M$ than $DM_n$ which would contradict our initial assumptions}. Since we assume that $w$ and its derivatives are finite, we know from section \ref{s12} that $M$ is then a sink and cannot be a saddle. Consequently, a transition from a saddle point $DM_n$ to a saddle point $M$ is impossible. In the same way one shows that a trajectory cannot go through a saddle point $M$ to a saddle point $DM_n$.\\
Last we show that if $DR_n$ is a saddle point, the next equilibrium point cannot be a saddle point of type $R$ and vice-versa. Indeed if $DR_n$ is a saddle, then in $DR_n$, $w=1/3$ and $\frac{dw}{dy_2}(0,y_{20})<0$. Following the continuous curve $(C_R)$, we thus arrive in $R$ with $w<1/3$ \footnote{Otherwise, $w$ would be equal to zero between $DR_n$ and $R$ on $(C_R)$ and there will be another $DR$ point closest to $R$ than $DR_n$ which would contradict our initial assumptions}. Once again, we derive from section \ref{s12} that $R$ is then a source and never a saddle, $w$ and its derivatives being finite. Hence, a transition from a saddle point $DR_n$ to a saddle point $R$ is impossible. In the same way one shows that a trajectory cannot go through a saddle point $R$ to a saddle point $DR_n$.\\
We conclude that when $w$ and its derivatives are finite, we cannot have two successive saddle epochs dominated by matter or radiation.
\bibliographystyle{unsrt}

\end{document}